\documentclass{article}    

\usepackage{graphicx}

\title{\textbf{Rule (4) and Continuous Observation}}  
\author{Richard Mould\footnote{Department of Physics and Astronomy, State University of New York, Stony Brook,
\mbox{New York} 11794-3800}}  
\date{}    
   
\begin{document}             

\maketitle              

\begin{abstract}

The effect of rule (4) on a series or parallel sequence of quantum mechanical steps is to insure that a conscious observer
 does not skip a step.  This rule effectively places the observer in continuous contact with the system.

\end{abstract}

\section*{Introduction}

 			A conscious observer cannot be continuously included in a quantum mechanical system without rule (4).  This is
demonstrated below in both series and parallel interactions.  If the Hamiltonian carries that system through a sequence of
discrete series or parallel states, then rule (4) will guarantee that a conscious observer who interacts with the system
will experience each step along the way.  No state is skipped over, so the observer is `continuously' present.  This
non-skip property of rule (4) is recognized in some previous papers \cite{RM1}-\cite{RM3}.  A summary of all four of
the governing rules and an explanation of their origin is found in ref.\ 3.

\section*{A Series Sequence}

Imagine a system given by
\begin{equation}
\Phi(t \ge t_0) = A_0\underline{B}_0 + A_1B_1 + A_2B_2 + A_3B_3 + \mbox{etc.}
\end{equation}
where the first component is initially normalized, and the others become \mbox{non-zero} after the interaction is
initiated at time $t_0$.  The Hamiltonian provides an interaction between the $0^{th}$ and $1^{st}$ components, the
$1^{st}$ and $2^{nd}$ components, the $2^{nd}$ and $3^{rd}$ components, etc.; however, there is no interaction term in the
Hamiltonian that skips over components.   That is, probability current will not flow into the $2^{nd}$ component until
the $1^{st}$ component has acquired some amplitude, and current will not flow into the $3^{rd}$ component until the
$2^{nd}$ has acquired some amplitude, etc.  

The letters $A$ in eq.\ 1 represent different states of a macroscopic apparatus, and the letters $B$ are understood to
represent brain states of the observer.  The initial brain state $\underline{B}_0$ is conscious as indicated by the
underline.  This means that the observer is directly conscious of the apparatus state $A_0$.  The remaining
\mbox{non-underlined} brain states in eq.\ 1 are not conscious.  They are called \emph{ready brain states} defined in
refs.\ 1-3, and are required by rule (2) to appear in eq.\ 1.  One of these ready states will become conscious the moment
it is stochastically chosen according to rules (1) and (3); and at that same time, all  the other components in eq.\ 1
will go to zero according to rule (3).  The first two rules provide for the existence of a stochastic trigger and ready
brain states, and the third provides for the collapse of the wave and the transfer of consciousness from $\underline{B}_0$
to the stochastically chosen brain component.  I will first discuss the interaction \emph{without} using rule (4).

As the interaction proceeds, probability current $J$ will flow from the $0^{th}$ component in eq.\ 1 to the $1^{st}$
component, and from there to the $2^{nd}$ component, etc.  Figure 1 shows a possible distribution of the square moduli of
these component at some time $t > t_0$.  

\begin{figure}[h]
\centering
\includegraphics[scale=1.2]{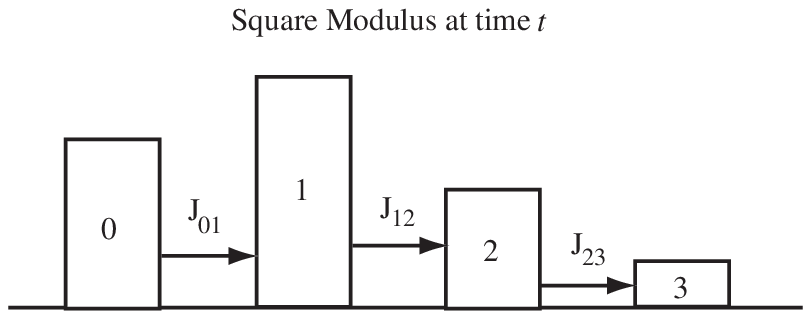}
\center{Figure 1}
\end{figure}

To give this a concrete interpretation, imagine that $A$ is a $\beta$-ray counter, where $A_0$ is the initial state  with
$0$ showing on the dial, $A_1$ is the counter with 1 on the dial, $A_2$ is the counter with 2 on the dial, etc.  Figure 1
shows the probability of each of these readings at time $t$, where the observer remains conscious of a zero reading up
to that time.  The observer will not become aware of counts 1, 2, or 3 until there is a stochastic hit on one of these
states as a result of the current flow into that state.  If the stochastic choice is the $2^{nd}$ component as a result of
current
$J_{12}$, then according to rule (3), there will be a state reduction and the system will immediately become
\begin{equation}
\Phi(t \ge t_{sc}) = A_2\underline{B}_2
\end{equation}
where $t_{sc}$ is the time of the stochastic choice.  If this happens, the observer will be unaware of the apparatus state
$A_1$ that has been skipped over.  His consciousness will go directly from $\underline{B}_0$ in eq.\ 1 to
$\underline{B}_2$ in eq.\ 2.  It is also possible that his consciousness will go directly to $\underline{B}_3$ as a result
of the current $J_{23}$ into that state.  

So far we have followed the first three rules, and it is obvious that they cannot guarantee that the observer is
continuously conscious of the apparatus.   Otherwise the observer would not skip over states.  This deficiency is remedied
by the addition of rule (4).

\vspace{.3cm}
\textbf{Rule (4)}: \emph{A transition between two components is forbidden if each is an entanglement containing a ready
brain state of the same observer.}
\vspace{.3cm}
 
This is a selection rule that forbids transitions that go between ready brain states of the same observer.  The rule was
initially adopted to avoid an anomaly that appears when a second observer makes a terminal observation of an interaction
that was witnessed from the beginning by a first observer (refs.\ 1 \mbox{and 3)}.  Without this rule, the second
observation might initiate a capture long after the first observer has established that the particle was not captured
during the interaction.  On this basis, the rule is necessary but seemingly ad hoc.  However, it has other important
consequences.

\section*{Continuous Consequences of Rule (4)}

When an observer is included in a $\beta$-counter system with rule (4) in effect, \mbox{eq.\ 1} becomes
\begin{equation}
\Phi(t_{sc} > t \ge t_0) = A_0\underline{B}_0 + A_1B_1 \hspace{.1cm}(+) \hspace{.1cm}A_2B_2\hspace{.1cm} (+)\hspace{.1cm}
A_3B_3 \hspace{.1cm}(+)\hspace{.1cm} \mbox{etc.}
\end{equation}
where the $0^{th}$ and $1^{st}$ components are the only ones that are actively involved before there is a stochastic hit at
time $t_{sc}$.  The parenthesis around the + sign means that current cannot flow between the indicated components because
each one contains a ready brain state, and rule (4) forbids a transition between ready brain states.  Therefore, current
cannot flow between the $1^{st}$ and $2^{nd}$ components in eq.\ 3, or between the $2^{nd}$ and $3^{rd}$ components, or
between any two of the higher order components.  So prior to a stochastic hit, current will only flow from the $0^{th}$
to the $1^{st}$ component as shown in fig.\ 2.  

\begin{figure}[h]
\centering
\includegraphics[scale=1.1]{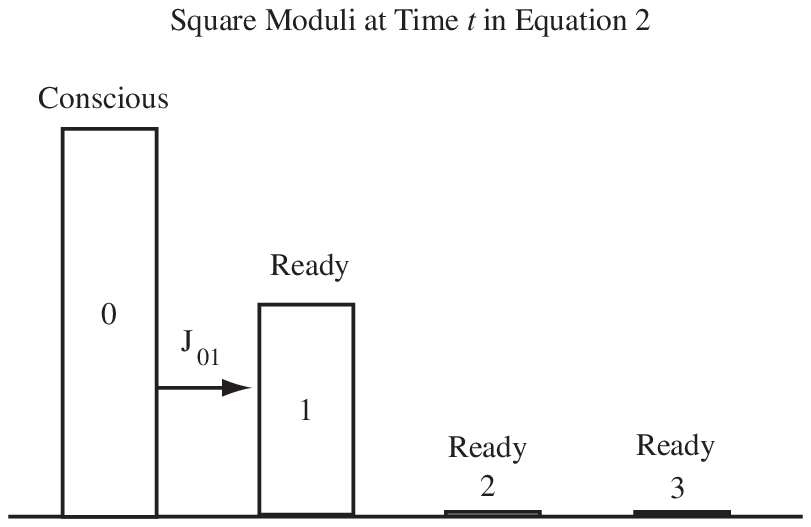}
\center{Figure 2}
\end{figure}

Therefore, with rule (4) in effect, the $1^{st}$ component \emph{will} be chosen and the correlated brain state
\emph{will} become conscious.  It will be chosen because all of the current from the normalized $0^{th}$ component will
pore into the $1^{st}$ component making $\int{J_{01}dt}= 1.0$, after which there will be a collapse to the $1^{st}$
component because of rule (3) and the associated brain state will become conscious.    Apparently, the  $1^{st}$
component will not be skipped over by the conscious observer.  

With the choice of the $1^{st}$ component, the process will begin all over again as shown in the middle diagram of fig.\
3.  This also leads with certainty to a stochastic choice and conscious awareness of the $2^{nd}$ component.  That
certainty is accomplished by the wording of rule (1).  This rule requires that the probability per unit time is given by
the current flow $J_{12}$ divided by the total square modulus at that moment.  The total integral $\int{J_{12}dt}$  is
less than 1.0 in the middle diagram of fig.\ 3, but it is restored to 1.0 when divided by the square modulus.   It
is therefore certain that the $2^{nd}$ component will be chosen and that the correlated brain state will become conscious. 
That is, the $2^{nd}$ component will not be skipped over by a conscious observer.

\begin{figure}[h]
\centering
\includegraphics[scale=0.9]{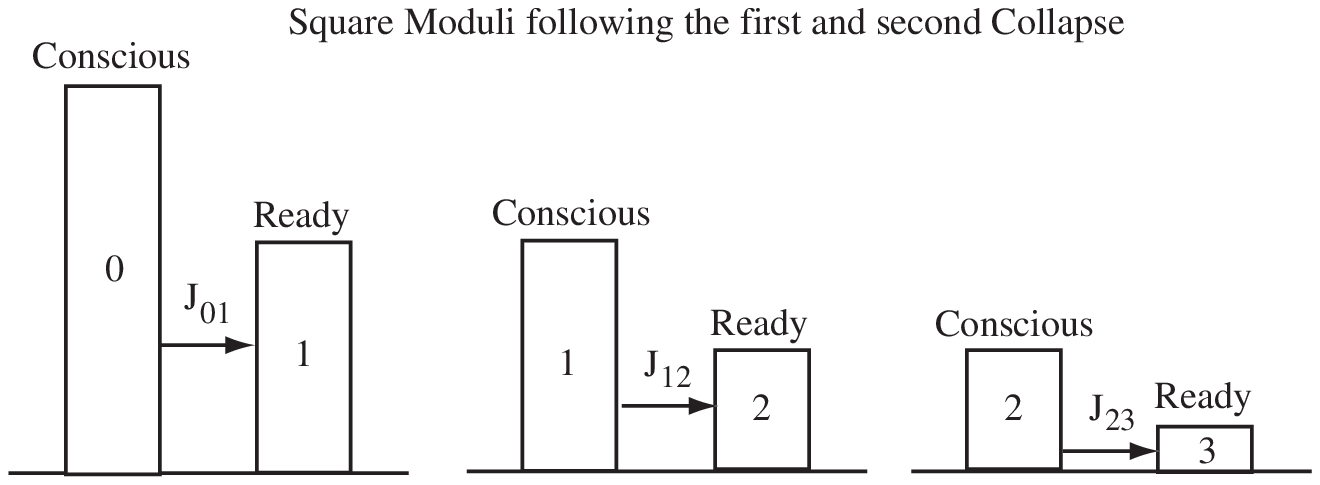}
\center{Figure 3}
\end{figure}

And finally, with the choice of the $2^{nd}$ component, the process will resume again as shown in the last diagram of
fig.\ 3.  This leads with certainty to a stochastic choice and conscious awareness of the $3^{rd}$ component.  

It follows that all of the counter states are experienced sequentially by a conscious observer.  Rule (4) has the effect
of placing the observer in continuous contact with the counter at every stage along the way.  This is in contrast to the
standard quantum mechanical observer who can only have momentary contact with a quantum mechanical system.  Even with a
high but finite frequency of momentary contacts as in the Zeno effect, there will always be a finite
probability that an intermittent observer will miss one of the counts on the counter \cite{MS}.  

It may seem that the effect of rule (4) on the dynamics of the system will alter the statistics.  This issue is dealt with
in another paper in which it is shown that rule (4) has \emph{no effect} on the over-all statistics 
\cite{RM4}.

\section*{A Parallel Sequence}

Now imagine a parallel sequence of states in which the process may go either clockwise or counterclockwise as shown in
fig.\ 4.  Each state includes a macroscopic piece of laboratory apparatus plus an observer who interacts with the
apparatus.  The Hamiltonian is assumed to provide clockwise interactions going from the $0^{th}$ to the $r^{th}$ state and
from the $r^{th}$ to the final state $f$, as well as a counterclockwise pathway from the $0^{th}$ to the $l^{th}$
state and from the $l^{th}$ to the final state $f$.  The Hamiltonian does not provide a direct route from the $0^{th}$ to
the final state.  The observer is initially conscious of the $0^{th}$ state as indicated by the underlined
$\underline{B}_0$.  

\begin{figure}[h]
\centering
\includegraphics[scale=1.1]{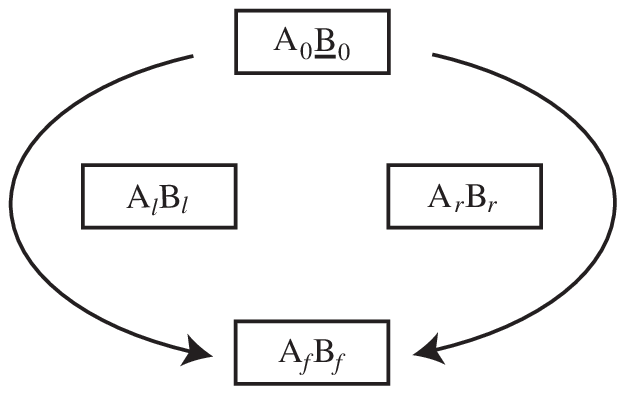}
\center{Figure 4}
\end{figure}

   	Without rule (4), current will initially flow from $A_0\underline{B}_0$ to the two intermediate states $A_rB_r$ and
$A_lB_l$, and as these states gain amplitude, current will flow from them to the final state.  Maybe the first stochastic
hit will result from the current flow from the $0^{th}$ state to one of the intermediate states.  In that case, either
$B_r$ or
$B_l$ will become conscious and all other states would be reduced to zero.  The process would then start over as current
flows from the surviving intermediate state to the final state.  This would surely lead to a stochastic hit on the
final state, causing $B_f$ to become conscious with all other components going to zero.  The observer would then see
either a complete clockwise or counterclockwise progression in which the intermediate state appears in proper sequence.  

	It is also possible that the first stochastic hit will result from the current flow into the final state from the
intermediate states.   In that case, $B_f$ will become conscious and the other states will go immediately to zero, causing
the observer to skip over the intermediate states entirely.  The observer will see the initial and the final state, but he
will not observe that the process went clockwise or counterclockwise.  The case is similar to that described by Heisenberg
of an elementary particle that is observed at an initial position and a final position, but nowhere in between.  Quantum
mechanics gives us no reason to believe that such a particle follows any particular path from the beginning to the end. 
Feynman showed that the particle exists in a superposition of all possible paths joining the initial to the final state. 
The same might be said of the intermediate states in fig.\ 4 when they are skipped over by the conscious observer.  In this
case, the state of the system between the initial and final observations would be a quantum mechanical superposition of
the two intermediate states \footnote{In this comparison, the difference between a macroscopic particle and a macroscopic
apparatus is: That environmental decoherence makes the components of the macroscopic superposition incoherent.}.   The
observer would have to conclude that it is intrinsically uncertain whether the apparatus followed a clockwise or a
counterclockwise path.

\section*{Rule (4) in Parallel Case}

With rule (4) in place, probability current cannot initially flow from either of the intermediate states to the final
state, for that would carry a ready brain state into another ready brain state of the same observer.  The dotted lines in
fig.\ 5 indicate these forbidden transitions.  A direct transition from the initial to the final state is also forbidden
because it is not provided for in the Hamiltonian.

\begin{figure}[h]
\centering
\includegraphics[scale=1.1]{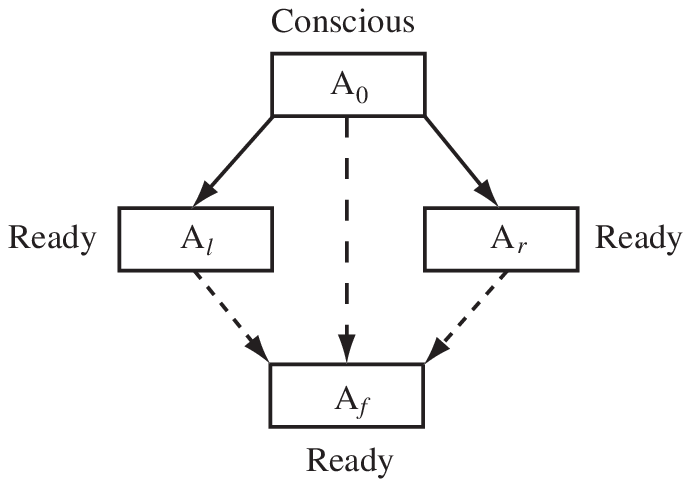}
\center{Figure 5}
\end{figure}

This means that the first stochastic hit \emph{will be} on the current flow from either the $0^{th}$ to the $r^{th}$ state,
or the $0^{th}$ to the $l^{th}$ state, thereby determining that the process will be clockwise or counterclockwise.   The
chosen intermediate state will then become conscious and all the other states will go to zero.  The process will then
resume with current flowing from the chosen intermediate state to the final state.  A second stochastic hit will make the
final state conscious with the intermediate state going to zero.  

It is therefore clear that with rule (4) in place, the observer will \emph{always} experience one or the other
intermediate state, so he will always know if the process is clockwise or counterclockwise.  He will never skip a step. 
Again,\mbox{rule (4)} assures that the observer's inclusion in the state of the system is a continuous one.

\section*{A Continuous Variable}
	The above examples involve the ``continuous" observation of a ``discontinuous" quantum mechanical variable in which the
stochastic chooser is required to go from one value of the variable to the next.  Rule (4) guarantees that none of these
finitely separated steps is passed over by an observer.  On the other hand, if the variable itself is classical and
continuous, then continuous observation is possible without the necessity of stochastic jumps.  In that case we do not
need rule (4) or any of the rules (1-4), for they do not prevent or in any way qualify classical motion.  

However, a classical variable may require a quantum mechanical jump-start.  For instance, the classical mechanical device
that is used to seal the fate of Schršdinger's cat begins its motion with a stochastic hit.  That is, the decision to
begin the motion (or not) is left to a $\beta$-decay.  The details of this are worked out in ref.\ 2 (eqs.\ 4 and 6 and
subsequent discussion).  It is shown there that rule (4) forces the motion to begin at the beginning, insuring that no
value of the classical variable is passed over in the presence of a conscious cat.  If the mechanical device is a
hammer that falls from a vertical position to release an anesthesia of some kind, rule (4) guarantees that the hammer will
begin its motion from the vertical position if the conscious cat is watching at the time.   Without rule (4), the cat might
see the hammer begin its fall at some other angle; because without it, probability current will flow
into angles other than zero.   However, with rule (4) in place, no angle will be passed over in the eyes of the on looking
cat.  

The more general relationship between quantum mechanical and classical change is described in another paper \cite{RM5}.

\end{document}